\documentstyle[aps,preprint,tighten]{revtex}
\begin{document}
\title{Scalar correlations in a quark plasma \\
       and low mass dilepton production}
\author{D. Blaschke, Yu. L. Kalinovsky  
\thanks{Permanent address: Laboratory for Computing Techniques and Automation, 
JINR Dubna, 141980 Moscow Region, Russia},
S. Schmidt} 
\address{Fachbereich Physik, Universit\"{a}t Rostock,
         D-18051 Rostock, Germany}
\author{H.-J. Schulze}
\address{Sezione INFN, Universit\`a di Catania, I-95129 Catania, Italy}
\maketitle

\begin{abstract}
We investigate possible consequences of resonant
scalar interactions for dilepton production from a quark plasma at the 
chiral phase transition.
It is found that this production mechanism is strongly suppressed
compared to the Born process and has no significance for present experiments.
 \\[2mm]
\end{abstract}
\date{today} 

\vskip 2cm

Recent experiments with ultrarelativistic S- and Pb-beams at the CERN-SPS have 
shown that dilepton production in the low mass region is strongly enhanced 
when compared with a simple extrapolation 
from proton-proton and proton-nucleus collisions \cite{ceres}.
An application of the previously developed standard approach which also 
includes a possible QCD phase transition \cite{classic} to the situation in
the CERES experiment \cite{sollfrank} for different sets of 
equations of state shows that 
(i) hadronic processes dominate the dilepton spectrum in the low mass region, 
but (ii) the experimental finding of low mass dilepton enhancement cannot be
reproduced within the standard scenario which neglects modifications of
either quark or hadron properties in a hot and dense medium.

Among possible explanations of the low mass dilepton enhancement the in-medium 
modification of the $\rho$ resonance is the standard one \cite{redlich,rho}. 
It is, however, debated whether chiral symmetry restoration influences the
fate of this vector resonance in the vicinity of the QCD phase transition 
\cite{rho}. In this context the r\^ole of critical phenomena related to 
scalar resonances should be considered.

Already in 1992, Weldon suggested that in a hot pion gas with finite chemical 
potential the scalar resonance can decay into a lepton pair \cite{weldon}.
This process is particularly interesting as a ``direct'' signal 
of chiral symmetry restoration in hot and dense matter, 
where the scalar and pseudoscalar mesonic modes become degenerate.  
With increasing temperature of the hadron gas, the hadronic decay channel 
($\sigma \rightarrow \pi\pi$) closes as soon as $m_\sigma < 2 m_{\pi}$ at 
$T\approx 150$ MeV \cite{hatsuda}. 
At this temperature the sigma meson could appear as a prominent 
resonance of the dilepton spectrum. 
However, the mechanism proposed by Weldon depends strongly on a pion 
chemical potential, and requires in particular a significant difference
of positive and negative pion densities, 
for which there is no experimental evidence.

In this article we rather scrutinize another mechanism in which scalar
correlations could be of importance, namely
we investigate whether resonant scalar $q \bar q$ interactions can be a 
source of dileptons in the mass region 
0.2 GeV $ \lesssim M_{e^+e^-} \lesssim $ 0.6 GeV.
The underlying physical scenario is the chiral symmetry restoration at the
phase transition of the hot meson gas to quark matter, whereby strong 
nonperturbative correlations of color-singlet $q \bar q$ pairs may persist
(``critical opalescence'' of quark matter \cite{huefner,rkb}).

We will show that due to the resonant interaction, 
the spectral density in the scalar channel of  
quark-antiquark annihilation in quark matter at the chiral phase transition 
might be enhanced by 1--2 orders of magnitude. 
The contribution of this process to the dilepton production rate,
however, is smaller than the perturbative thermal Born process and cannot be
considered as a source for the dilepton enhancement observed by the CERES 
collaboration.

Due to the built-in chiral symmetry it seems reliable to use in this 
report the simplest SU(2) version of the Nambu--Jona-Lasinio 
(NJL) model at finite temperature and chemical potential in order to study 
the scalar correlations in the quark phase \cite{sbk} and to give at least 
order of magnitude estimates of the physical phenomena we are interested in.

We consider the production (by $q\bar q$ annihilation) 
of a virtual photon, i.e.~lepton pair,
with invariant mass $M^2=(p_1+p_2)^2$ and three-momentum 
$\rm P= |{\bf p_1}+{\bf p_2}|$ in a locally 
thermalized medium that is characterized by a fluid four-velocity $u$. 
The temperature $T$ and quark chemical potential $\mu$ 
in a given fluid cell determine uniquely the 
constituent quark mass $m$ as solution of the gap equation \cite{sbk}.
The lepton mass is denoted by $m_l$.

The spin-averaged annihilation cross section in lowest order 
of the electromagnetic interaction \cite{classic} is given by the 
general expression
\begin{eqnarray} 
  \sigma\left[ q(p_1) \bar{q}(p_2) \rightarrow \gamma(M,{\rm P}) 
               \rightarrow l^+ l^-\right] &=&
  {\alpha\over3 M^4}  
  { L(M)H(M,{\rm P})\over \sqrt{1-{4m^2/M^2}}} ~,
\label{e:wq}
\\ 
  L(M) &=& \left( 1+ {2m_l^2\over M^2} \right) 
  \sqrt{1-{4m_l^2\over M^2}} \theta(M^2-4m_l^2) ~,
\end{eqnarray}
where $H=H_\mu^\mu$ corresponds to the contraction of the so-called
hadronic tensor
\begin{equation}
  H_{\mu\nu} = \overline{\sum_{\rm spin}}
  \big\langle q(p_1)\bar{q}(p_2) \big| J_\mu(0) \big| 0 \big\rangle 
  \langle 0| J_\nu(0) | q(p_1)\bar{q}(p_2) \rangle ~. 
\end{equation}

The inclusion of scalar correlations corresponds to the following 
modification of the electromagnetic quark current operator 
with respect to the Born process:
\begin{eqnarray}
  \langle q(p_1)\bar{q}(p_2) | J_\mu(0) | 0 \rangle &=&
   \left[ \bar{v}(p_1) e_q \gamma_\mu u(p_2) \right]
\rightarrow
  \left[ \bar{v}(p_1) u(p_2) \right]
  \frac{K}{1 - J(M,{\rm P})}  I_\mu (M,{\rm P}) ~.
\label{matrel2}
\end{eqnarray}
Here the polarization operator $J(M,{\rm P})$ 
for the scalar isoscalar channel at finite temperature and chemical potential
can be evaluated in the NJL model by using the standard techniques of
finite temperature field theory \cite{tft} with the result
\begin{eqnarray}
  J(M,{\rm P}) &=&
  i K \int \frac{d^4k}{(2\pi)^4}
  \mbox{Tr} \left[ G(k) G(k-P) \right] 
\nonumber 
\\\label{loopint}
  &=&  \frac{N_C N_F K}{\pi^2}
  \int_0^\Lambda \frac{ {\rm k}^2 d{\rm k} }{\omega}
  \frac{\mbox{sinh}(\omega/T)} 
       {\mbox{cosh}(\omega/T) + \mbox{cosh}(\mu/T) }
  \left[ 1- \frac{M^2-4m^2}{8{\rm Pk}} \mbox{ln}(F_+ F_-) \right]~,
\end{eqnarray}
where $\omega=\sqrt{{\rm k}^2+m^2}$ and $E=\sqrt{{\rm P}^2+M^2}$ are the quark
and photon energies, respectively, and 
\begin{equation}
F_{\pm} =
  \frac {M^2 \pm 2E\omega +2{\rm Pk}}
        {M^2 \pm 2E\omega -2{\rm Pk}} ~.
\end{equation}
The loop integral (\ref{loopint}) has an imaginary part
\begin{eqnarray}\label{loopimag}
{\rm Im} J(M,{\rm P}) = -\frac{N_C N_F K T}{8\pi {\rm P}}(M^2-4m^2)
\ln\bigg[\frac{\cosh(\omega_{{\rm max}}/T)+\cosh(\mu/T)}
{\cosh(\omega_{{\rm min}}/T)+\cosh(\mu/T)}\bigg]\,\,,
\end{eqnarray} 
with $\omega_{{\rm max,min}}=\big[E\pm{\rm P}\sqrt{1-4m^2/M^2}\big]/2$.
The imaginary part (\ref{loopimag}) is non-vanishing for $M>2m$ and  
corresponds to the decay width of the scalar meson in the ${\bar q}q$- channel.
Note that we use a simple version of the NJL model with the scalar 
coupling constant $K$ and a three-momentum cutoff $\Lambda$.

The loop integral describing the transition
$\sigma \rightarrow \gamma$ has the form
($Q$ is the charge operator for the quarks)
\begin{eqnarray}
  I_\mu(M,{\rm P}) &=&
  \int \frac{d^4k}{(2\pi)^4} \mbox{Tr}
  \left[ G(k) \gamma_\mu Q G(k-P) \right] 
\nonumber \\
  &=& \frac{4}{3} N_C e m \int \frac{d^4k}{(2\pi)^4}
  \frac{2 k_\mu - P_\mu} {(k^2-m^2)[(k-P)^2-m^2 ]}~.
\label{int1}
\end{eqnarray}
Using the condition of charge conservation, $P_\mu I^\mu =0$,
one can derive the useful relation
\begin{eqnarray}
  I_\mu I^\mu = -\frac{M^2}{{\rm P}^2} (I_\mu u^\mu)^2  ~,
\end{eqnarray}
and it is then practical to define the dimensionless quantity
\begin{eqnarray}
  \widetilde{I} = {I_\mu u^\mu \over 4em{\rm P}}
  &=& {1\over {\rm P}} \int \frac{d^4k}{(2\pi)^4}
  \frac{2k_0-E}{(k^2-m^2)[(k-P)^2-m^2]} 
\nonumber \\
  &=& \frac{1}{(4\pi {\rm P})^2}\int_m^\infty d\omega
  \,\delta n(\omega) 
  \left[ (2\omega+E) \ln(F_+) + (2\omega-E) \ln(F_-) \right]  
\label{e:i}
\end{eqnarray}
with
\begin{equation}
  \delta n(\omega) = \frac{\sinh(\mu/T)}
  {\mbox{cosh}(\omega/T)+\mbox{cosh}(\mu/T)} ~. 
\end{equation}

This loop integral does only give 
a nonvanishing contribution if two conditions are fulfilled:
(a) The difference $\delta n$ of particle and antiparticle distributions 
is nonzero because of a chemical potential corresponding to a 
finite baryon number.
(b) The three-momentum $\rm P$ of the pair must be nonzero 
in the rest system of the medium (fluid) in which temperature and 
chemical potential are defined.

Coming back to Eq.~(\ref{e:wq}), we obtain
\begin{eqnarray} 
  H_{\rm Born}(M) &=& e_q^2 (M^2+2m^2) ~,
\\
  H_{\rm Resonance}(M,{\rm P}) &=& 8 e^2 (M^2-4m^2)
  \left| D(M,{\rm P})\widetilde{I}(M,{\rm P}) \right|^2 ~,
\end{eqnarray}
where the dimensionless quantity
\begin{equation}
  D(M,{\rm P}) =
  \frac{K m M }{1-J(M,{\rm P})}
\label{e:d}
\end{equation}
is related to the propagator of the scalar $q\bar{q}$ correlation.

The final result for the resonance cross section for dilepton production 
relative to the Born cross section is
\begin{equation}
  \frac{\sigma_{\rm Resonance} (M,{\rm P})}{\sigma_{\rm Born}(M)} =
  8\frac{\alpha}{\alpha_q} 
  \left| D(M,{\rm P}) I(M,{\rm P}) \right|^2 
\label{e:rat}
\end{equation}
with 
$I(M,{\rm P}) = \sqrt{(M^2-4m^2)/(M^2+2m^2)} \widetilde{I}(M,{\rm P})$.

For the numerical calculation we use the 
 NJL parameters fixed as in Ref.~\cite{b+96} 
\footnote{$K=9.45\;\rm GeV^{-2}$, $\Lambda=660\;\rm MeV$, 
current quark mass $m_0=5.35$~MeV}.
We choose two representative sets of temperature and 
(quark) chemical potential and obtain
(all numbers are in MeV):
(a) $T=170$, $\mu=80$, $m=150$, $m_\sigma=330$; and 
(b) $T=240$, $\mu=110$, $m=33$, $m_\sigma=340$.
The first set is motivated by a recent fit of thermodynamical parameters 
to experimental hadron abundancies \cite{stachel}.
It yields a sigma mass close to the double quark mass threshold, and 
consequently the sigma propagator $D$ is strongly peaked.
While the first set (a) corresponds to a scenario in the vicinity of the chiral
phase transition, in the second set (b) we have chosen rather extreme 
conditions were the chiral symmetry is almost restored
and results in a much smaller constituent quark mass, and a larger width of
the sigma resonance.

The results of the calculation are given in Figs.~1 and 2, where we 
plot the moduli of the transition integral $I$, Eqs.~(\ref{e:i},\ref{e:rat}), 
and the scalar propagator $D$, Eq.~(\ref{e:d}), 
as well as the resulting ratio of the resonance and the Born cross section, 
Eq.~(\ref{e:rat}), where we have
carried out an isospin average $\alpha/\alpha_q\rightarrow 18/5$.
It shows that there is an enhancement of the scalar propagator 
$D(M,{\rm P})$ which, however, 
cannot overcome the smallness of the transition function $I$.
The resulting cross section is always suppressed by at least two 
orders of magnitude relative to the Born process, even for the most
favorable kinematical conditions.

It can be concluded that the proposed sigma-induced dilepton 
production process has no relevance for present-day experiments,
where even the thermal Born $q\bar q$ cross section    
is negligible relative to hadronic decay contributions in the low mass 
region of dilepton production.

\begin{figure}
\caption[]{
Dependence on invariant mass $M$ and three momentum $\rm P$ of
the transition integral $|I|$ (top figure), 
the scalar propagator $|D|$ (middle figure),
and the ratio of resonance and Born cross section (bottom figure)
for a temperature $T=170$~MeV and quark chemical potential $\mu=80$~MeV.
}
\end{figure}

\begin{figure}
\caption[]{
Same as Fig.~1, but for $T=240$~MeV and $\mu=110$~MeV.
}
\end{figure}


\begin{references}
\bibitem{ceres}
G. Agakichiev et al., Phys. Rev. Lett. {\bf 75} (1995) 1272;\\
A. Drees (CERES Collaboration), Nucl. Phys. {\bf A 610} 536c.
\bibitem{classic}
L.D. McLerran and T. Toimela, Phys. Rev. {\bf D 31} (1985) 545;\\
K. Kajantie, J. Kapusta, L. McLerran and A. Mekjian, Phys. Rev. {\bf D 34} 
(1986) 2746;\\
J. Cleymans, J. Fingberg and K. Redlich, Phys. Rev. {\bf D 35} (1987) 2153;\\ 
K. Kajantie and P.V. Ruuskanen, Z. Phys. {\bf C 44} (1989) 167. 
\bibitem{sollfrank}
J. Sollfrank, P. Huovinen, M. Kataja, P.V. Ruuskanen, M. Prakash and
R. Venugopalan, Phys. Rev. {\bf C 55} (1997) 392.
\bibitem{redlich}
F. Karsch, K. Redlich and L. Turko, Z. Phys. {\bf C 60} (1993) 519.  
\bibitem{rho}
G.Q. Li, C.M. Ko and G. Brown, Phys. Rev. Lett. {\bf 75} (1995) 4007;\\
G.Q. Li, C.M. Ko, G.E. Brown and H. Sorge, 
Nucl. Phys. {\bf A 611} (1996) 539;\\
G. Chanfray, R. Rapp and J. Wambach, Phys. Rev. Lett. {\bf 76} (1996) 368;
Nucl. Phys. {\bf A 617} (1997) 472.
\bibitem{weldon}
H.A. Weldon, Phys. Lett. {\bf B 274} (1992) 133.
\bibitem{hatsuda}
T. Hatsuda and K. Kunihiro, Phys. Lett. {\bf B 185} (1987) 304.
\bibitem{huefner}
J. H\"ufner, S.P. Klevansky and P. Rehberg, Nucl. Phys. {\bf A 606} (1996) 260.
\bibitem{rkb}
P. Rehberg, Yu. Kalinovsky and D. Blaschke, 
{\it Critical Scattering and Two-Photon Spectra for a Quark/Meson Plasma }
Nucl. Phys. {\bf A} (1997), in press. 
\bibitem{sb}
H.-J. Schulze and D. Blaschke,
Phys. Lett. {\bf B 386} (1996) 429.
\bibitem{sbk}
S. Schmidt, D. Blaschke and Yu.L. Kalinovsky, 
Phys. Rev. {\bf C 50} (1994) 435, and references therein.
\bibitem{tft}
J.I. Kapusta, {\em Finite-Temperature Field Theory} 
(Cambridge University Press, Cambridge, England, 1989);\\
M. Le Bellac, {\em Thermal Field Theory} 
(Cambridge University Press, Cambridge, England, 1996).
\bibitem{b+96}
D. Blaschke, Yu.L. Kalinovsky, G. R\"opke, S. Schmidt and M.K. Volkov,
Phys. Rev. {\bf C 53} (1996) 2394;\\
S. Schmidt, Dissertation, Rostock 1995, unpublished.
\bibitem{stachel}
P. Braun-Munzinger, J. Stachel, J.P. Wessels and N. Xu, Phys. Lett. {\bf B 365}
(1996) 1.
\end{references}
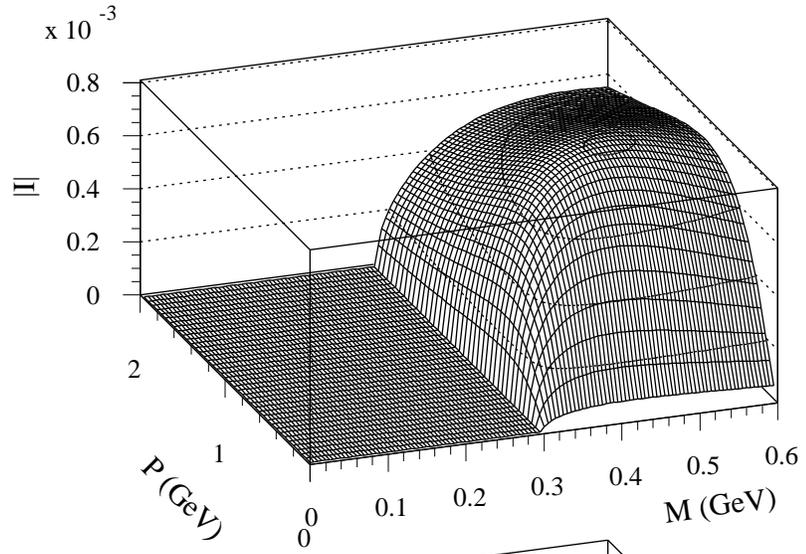
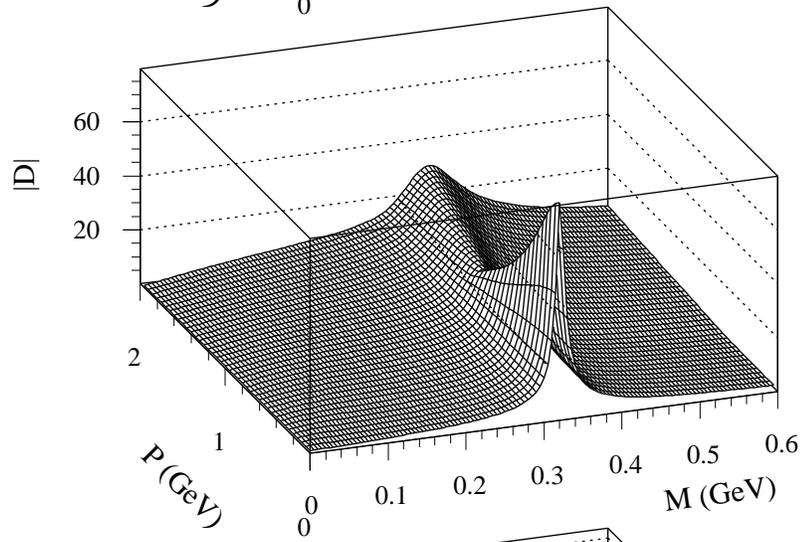
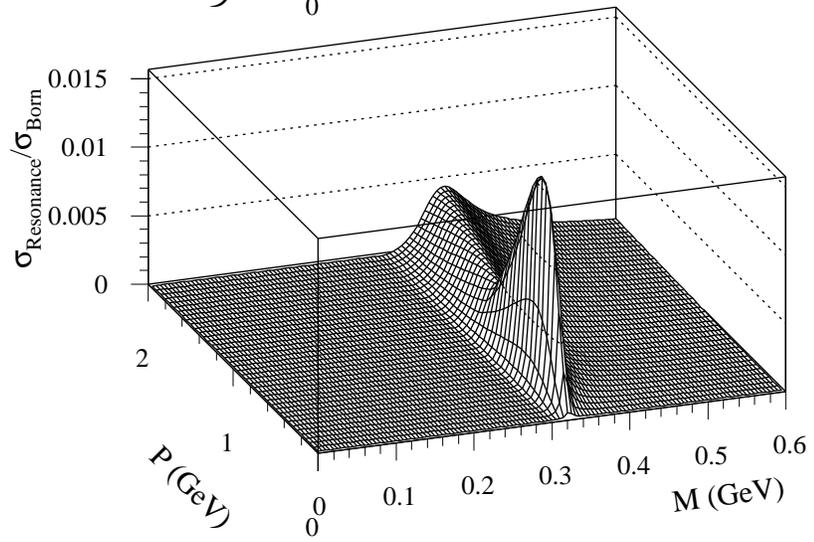
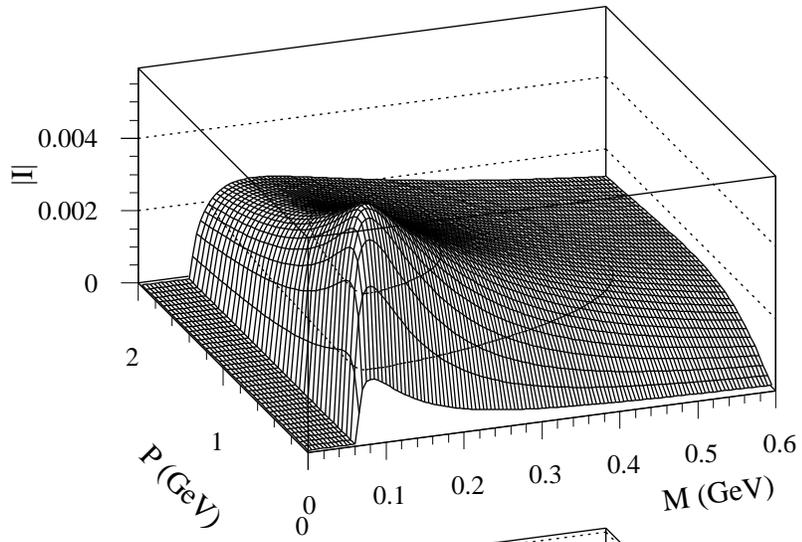
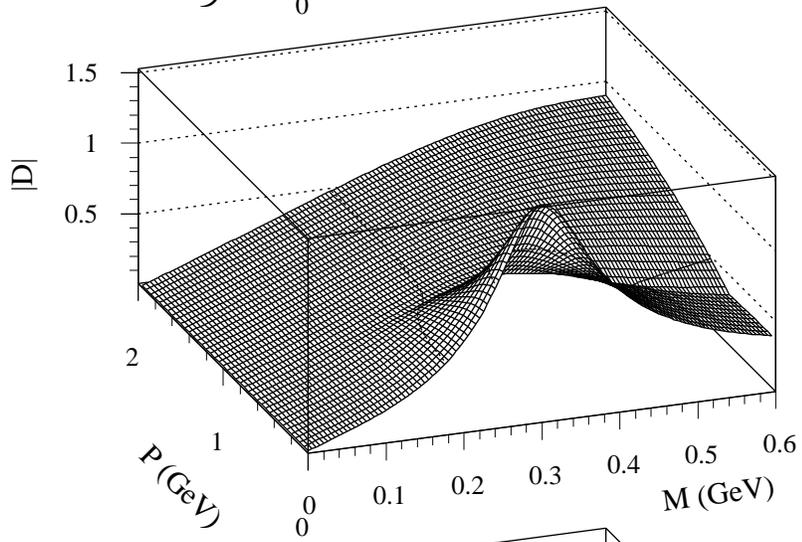
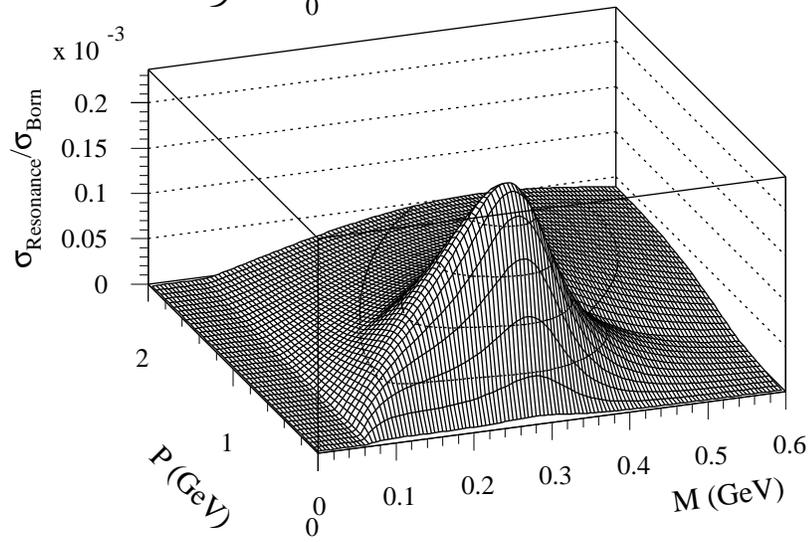
\end{document}